# Preparation of high-yield and ultra-pure $Au_{25}$ nanoclusters: towards their implementation in real-world applications

Michael Galchenko[1], Raphael Schuster[2], Andres Black[1], Maria Riedner[2], Christian Klinke[1,3,*]

*1 Institute of Physical Chemistry, University of Hamburg,*
*Martin-Luther-King-Platz 6, 20146 Hamburg, Germany*

*2 Institute of Organic Chemistry, University of Hamburg,*
*Martin-Luther-King-Platz 6, 20146 Hamburg, Germany*

*3 Department of Chemistry, Swansea University – Singleton Park,*
*Swansea SA2 8PP, United Kingdom*

**Abstract**

Colloidal approaches allow for the synthesis of Au nanoclusters (NCs) with atomic precision and sizes ranging from a few to hundreds of atoms. In most of the cases, these processes involve a common strategy of thiol etching of initially polydisperse Au nanoparticles into atomically precise NCs, resulting in the release of Au-thiolate complexes as byproducts. To the best of our knowledge, neither the removal of these byproducts nor the mass spectra in the relevant mass region were shown in previous studies. A thorough analysis of inorganic byproducts in the synthesis of $Au_{25}$ NCs reveals that published protocols lead to $Au_{25}$ NCs in vanishingly small quantities compared to their byproducts. Three purification methods are presented to separate byproducts from the desired $Au_{25}$ NCs which are proposed to be applicable to other promising Au NC systems. Additionally, critical factors for a successful synthesis of $Au_{25}$ NCs are identified and discussed including the role of residual water. An important finding is that the etching duration is very critical and must be monitored by UV-Vis spectroscopy resulting in synthetic yields as high as 40%.

**Introduction**

Among metals, Au particles surrounded by an organic monolayer (ligands) is the most studied system which is not surprising considering that their first colloidal synthesis dates back to 1857.[1] The transition between metal nanoparticles (NPs) with a continuous energy band structure and molecules with discrete energy states is embodied by metal nanoclusters (NCs) consisting of few to few-hundred atoms. Contrary to their larger counterparts, Au NCs can be synthesized in an atomically precise fashion and experience strong quantum confinement.[2] In this size regime unique properties appear, such as photoluminescence with quantum yields up to 40%, (photo-)catalytic activity or remarkable thermal and colloidal stability.[3,4] Their ultrasmall size provides Au NCs with several features for potential applications beyond fundamental research: Due to their high specific surface area and fraction of low-coordinated atoms, they are considered to have potential in heterogeneous catalysis,[5,6] and highly luminescent Au NCs have found tremendous application in nanomedicine.[7,8]

Various approaches are known for synthesizing atomically precise Au particles containing up to 333 Au atoms covered with weakly or strongly bound ligands,[9] as there are triphenylphosphine ($PPh_3$) and phenylethanethiol (PET), respectively. Most of these involve a one-step (e.g. for $[Au_{11}(PPh_3)_8Cl_2]^+$ or $[Au_{25}(PET)_{18}]^+$)[10,11] or two step (e.g. for $[Au_{38}(PET)_{24}]$, $[Au_{25}(TPP)_{10}(PET)_5X_2]^{2+}$ or $Au_{37}(TPP)_{10}(PET)_{10}X_2]^+$)[12-14] synthesis and a common strategy of thiol etching. During this process, thiol molecules release gold atoms from polydisperse Au NPs to create the desired atomically precise Au NCs. The type of byproducts formed and their removal with the proposed purification protocol remained unclear, since, to the best of our knowledge, mass spectra are never shown in the mass range where few Au-atom containing complexes give rise to a signal. However, this distinct mass range contains the signal at m/z 1055 of the $[Au_2TPP_2PET]^+$ complex which appeared to be a major byproduct in the synthesis of $[Au_{25}(TPP)_{10}(PET)_5X_2]^{2+}$ NCs.[15] In the case of purely thiol stabilized Au NCs, a thiol etching mechanism is proposed involving a Au(I)-thiolate monomer as a leaving group.[16] Indeed, removal of byproducts including few Au-atom comprising and charged complexes may be relatively easy in the case of uncharged Au NCs as the main products. However, due to the similar solubility of product and byproducts, the separation becomes more challenging for charged NCs, such as the highly fluorescent $[Ag_xAu_{25-x}(PPh_3)_{10}(SR)_5Cl_2]^{2+}$, the electron transfer catalyst $[Au_{25}(PET)_{18}]^+$ or the herein studied NC.[3,17] In order to implement these Au NCs successfully in an application, purity in terms of organic and inorganic byproducts is indispensable, but usually not covered sufficiently or not at all in the literature. The importance of purity becomes evident thinking of small gold-containing byproducts impacting the catalytic properties of the Au NC system.[18] Most substantially, such impurities can cause quenching of catalytic activity or catalyst poisoning and explain the poor catalytic activity of specific Au NCs compared to others.[20] Furthermore, small gold-containing byproducts can undesirably bioconjugated[19] to a target instead of the Au NC or quench the NCs fluorescence.

This work thoroughly analyzes the inorganic byproducts in the $Au_{25}$ NC synthesis and reports the discovery of a previously unknown Au complex. This was achieved by high-resolution mass spectrometry (MS) covering the low m/z range and an established HPLC method, which is suggested to be applicable to a preparative separation of the above mentioned charged NCs. A hitherto unparalleled analysis by means of MS combined with UV-Vis spectroscopy was applied suggesting that all so far published protocols are mainly leading to $[Au_2TPP_2PET]^+$ complexes and $Au_{25}$ NCs only in vanishingly small quantities. Finally, two methods for the separation of $Au_{25}$ NCs from its byproducts are presented. In addition, potentially important key factors for a successful synthesis of $Au_{25}$ NCs are discussed, including an optimized reaction time, the role of residual water and the relationship between these two factors.

**Experimental**

**Synthesis of [Au$_{25}$(PPh$_3$)$_{10}$(PET)$_5$X$_2$]$^{2+}$**

Au$_{25}$ NCs with the formula [Au$_{25}$(PPh$_3$)$_{10}$(PET)$_5$X$_2$]$^{2+}$ (X= Cl or Br) were synthesized as follows. HAuCl$_4$ · 3H$_2$O (0.100 g, dissolved in 3 mL of Millipore H$_2$O) was added into a 8 mL toluene solution of tetraoctylammonium bromide (TOAB, 0.145 g) and stirred vigorously (400 rpm) at room temperature for 15 min. Indicating the complete phase transfer of the gold compound from aqueous to toluene phase, the aqueous phase became colorless and clear and was removed by a pipette. The following steps appeared to be important for a water-free transfer of the gold compound: First, the organic phase was carefully transferred to a different glass ware with several glass pipettes, which were exchanged upon appearance of water-droplets sticking to the glass. The organic phase subsequently was poured into the 3-neck flask leaving further water-droplets behind. Subsequently, Triphenylphosphine (PPh$_3$, 0.180 g) was added into the flask under stirring (400 rpm). Within few to tens of seconds, the solution became cloudy white. If this happens delayed water residues are present in the suspension. After 15 min freshly dissolved sodium borhydride (NaBH$_4$, 0.026 g, dissolved by ultrasonication in 5 mL of EtOH) was injected rapidly to reduce Au$^I$(PPH$_3$)X (starting material) to Au NPs. After 2 hours, the dispersion was dried by rotary evaporation at 50°C, resulting in a dry black solid. White solid residues in the rotary evaporator flask indicate a delayed formation of the starting material (discussed in the results). The black solid was mixed with 20 mL dichlormethane (DCM), vortexed and centrifuged at 16000 rcf for 3 min. The resulting black supernatant was then heated to 40°C under reflux. Phenylethanthiol (PET, 300 uL) was added to the black dispersion, which was stirred at 400 rpm.

When the UV-Vis optical spectrum has evolved as shown in Fig. 1A or latest after 96 h, the yellow/brownish dispersion was dried by rotary evaporation at 50°C. An oily black product was obtained and transferred by means of 2 mL DCM into a glass centrifuge flask.

**Purification of $[Au_{25}(PPh_3)_{10}(PET)_5X_2]^{2+}$**

Subsequently, the suspension was precipitated with 80 mL of hexane, centrifuged at 3000 rcf and the supernatant was removed. The DCM/Hexane washing step was repeated four more times ($V_{DCM}/V_{Hex}$= 1/20; 1/40; 0,5/40; 0,5/40). Of note is that $Au_{25}$ NCs were still stable even after eight repetitions, which indicates a colloidal stability which is different to their larger counterparts. Finally, the $[Au_{25}(PPh_3)_{10}(PET)_5X_2]^{2+}$ were extracted by mixing with 10 mL MeOH, vortexing and centrifugation at 16000 rcf. The synthesis could easily be scaled up twofold by doubling starting and subsequently added materials mass/volume as well as purification steps.

Selective precipitation was performed as follows. 1 mL of the "purified" $Au_{25}$ NC suspension was dried by vacuum in a 10 mL glass vial. Of note is that during drying the vial was rotated. Thereby, the suspension was spread over a large area being beneficial for the purification process. Subsequently, a mixture of methanol/water (Millipore) (10 mL, 20/80 vol%) was added to the dry material. The brownish/black material remained undissolved and the supernatant colorless or slightly yellowish. The supernatant was discarded and the brownish/black solid could be redissolved in a residue-free fashion in 3mL of methanol.

Alternatively, $Au_{25}$ NCs can be crystallized with a modified protocol based on a known procedure[15]. Briefly, 10 mL of the "purified" $Au_{25}$ NC suspension (optical density = 9) was mixed with excess $NaSbF_6$ (240 mg) and stirred for 15 min in DCM. After removing residual $NaSbF_6$ (white solid) by means of centrifugation, the suspension was transferred to an unsealed 20 mL glass or Teflon vial. Subsequently, this vial was placed in a glass container, equipped with a ground glass joint. After filling hexane into the glass container until it reached a similar level as DCM, the container was left sealed in the fridge. After 5-7 days the suspension became colorless and clear leading to solid residues on glass walls and black, needle-like crystals. Further purification, like mixing with hexane and discarding it, was not applied since otherwise crystals tended to stick to the glass walls. The crystals can be transferred to a substrate for subsequent studies using a pipette filled with hexane or water or alternatively redissolved in methanol.

The following chemicals were used as received. Triphenylphosphin (99%), ethanol (99.98%) and toluene (99.85%), were purchased from Acros Organics. Methanol (99,8%) and n-hexane (96%) were purchased from VWR. Gold(III) chloride trihydrate (99,995%) and tetraoctylammonium bromide (98%) were obtained from Alfa Aesar. Dichlormethane (DCM, 99,5%) was obtained from Grüssing. Water was purified using a Millipore-Q System (18.2 MΩ cm). UV/Vis extinction spectra were obtained with a Cary 50 spectrophotometer. Transmission electron microscopy (TEM) images were obtained using Philips CM 300. High-resolution TEM images were obtained using a JEOL JEM 2200 FS. All the TEM samples were prepared by dropping 10 mL of diluted methanol suspension onto carbon-coated TEM grids followed by solvent evaporation under ambient conditions.

**Electrospray ionization quadrupole time-of-flight mass spectrometry**

5 µL of the sample (optical density 1 in ACN), diluted further 1:5 using dH$_2$O / Acetonitrile (CAN, HPLC grade) 50/50 v/v, were injected via direct infusion into an ESI-Q-TOF mass spectrometer (maXis, Bruker Daltonik) coupled to an UHPLC (Dionex Ultimate 3000, Thermo Scientific). As an eluent ddH$_2$O + 0.1% formic acid (FA, HPLC grade) / ACN + 0.1 % FA 50/50 v/v was used with a flow rate of 200 µL/min. MS-spectra were recorded in positive ion mode with a mass range of 800-5000 Da and a spectra rate of 1 Hz. The capillary voltage was set to 4500 V, the source temperature to 200 °C, the nebulizer pressure to 3.0 bar and the drying gas to 8.0 L/min.

**Liquid chromatography electrospray ionization quadrupole time-of-flight mass spectrometry**

5 µL of the same sample were injected into an UHPLC system (Dionex Ultimate 3000, Thermo Scientific) coupled to an ESI-Q-TOF mass spectrometer (maXis, Bruker Daltonik). For separation the compounds were loaded onto a reversed phase separation column (Extend C18, 1.8 µM, 2.1 x 50 mm, Agilent Technologies, Santa Clara, USA) and a linear gradient with ddH$_2$O + 0.1 % FA (buffer A) and ACN + 0.1 % FA (buffer B) as mobile phase and a flow rate of 300 µL/min was applied. In the beginning the concentration of buffer B was set to 50 % and was increased to 95 % within 30 minutes, kept constant for two minutes, was decreased to 50 % within 3 minutes and finally kept constant for five minutes. MS spectra were recorded in positive ion mode with a mass range of 800-5000 Da and a spectra rate of 1 Hz. The capillary voltage was set to 4500 V, the source temperature to 200 °C, the nebulizer pressure to 3.0 bar and the drying gas flow to 8.0 L/min.

**Matrix-assisted laser desorption/ionization time-of-flight mass spectrometry**

5 µL of the same sample were mixed with 1 µL of MALDI matrix (2,5-Dihydroxybenzoic acid, 20 mg/mL in $dH_2O$/ACN 70/30 v/v and 0.1 % TFA) and the mixture was spotted and dried on a MALDI groundsteel target. The target was injected into a MALDI-TOF-TOF mass spectrometer (UltrafleXtreme, Bruker Daltonik) and MS spectra were recorded in positive ion mode with a mass range of 500-5040 Da, a laser intensity of 27 % and 1000 laser shots per spectrum in partial sample mode. Ion source 1 voltage was set to 20 kV, ion source 2 voltage to 17.85 kV, lens voltage to 8.8 kV, reflector 1 voltage to 21 kV and reflector 2 voltage to 11 kV.

To obtain MS/MS spectra the ions of interest were isolated and fragmented using MALDI-LIFT-MS/MS. MS/MS spectra were recorded in positive ion mode, with a laser intensity of 29-35 % and 9000 laser shots per spectrum in partial sample mode. Ion source 1 voltage was set to 7.5 kV, ion source 2 voltage to 6.85 kV, lens voltage to 3.5 kV, reflector 1 voltage to 29.5 kV, reflector 2 voltage to 13.85 kV, LIFT 1 voltage to 19.00 kV and LIFT 2 voltage to 3.2 kV.

**Results and discussion**

On the basis of known procedures[13,15] a modified approach for the synthesis of $Au_{25}$ NCs with the formula $[Au_{25}(TPP)_{10}(PET)_5X_2]^{2+}$ was established. Initially, polydisperse 1-3 nm sized $Au_n$ TPP-stabilized NPs were synthesized, followed by size focusing through thiol etching, which produced the desired $Au_{25}$ NC. Finally, ultra-pure $Au_{25}$ NCs were obtained by separation from major inorganic byproducts. Fig. 1A shows UV-Vis optical spectra of aliquots taken at different times during the thiol-etching step, showing how large, polydisperse $Au_n$ NPs are converted to $Au_{25}$ NCs (crystal structure shown in Fig. 1A inset). Before size focusing, the rather featureless optical spectrum is typical of $Au_n$ NPs smaller than 2.5 nm.[22] As the particle size distribution narrows over time, distinct and gradually sharpening peaks appear. The peak at 670 nm corresponds to the HOMO to LUMO transition with a wave function localization around the vertex position. Higher excitations are partially due to HOMO-$n$ to LUMO+$n$ transitions (n>0) in the $Au_{13}$ sub-units comprising the $Au_{25}$ NC.[14]

The increasing intensity of these peaks serves as convenient markers to monitor the etching process. Plotted in Figure 1B is the height of both peaks as a function of etching duration, which respectively were measured from the peak maximum to minimum at shorter wavelength. It becomes evident that after an initial decrease, the peak height at 415 nm increases and levels off at an extinction value between 0.07 and 0.09 (100 uL withdrawn aliquot diluted to 3 mL). This is the highest possible extinction value demonstrating the termination of the size focusing process. Furthermore, as indicated by the error bars, the necessary etching duration can fluctuate and should be monitored by UV-VIS spectroscopy.

We propose residual water accidently transferred during the phase transfer of the Au compound from water to the organic phase to be responsible for that. As an increased amount of water was observed to prevent or delay the white clouding upon TPP addition and thus AuTPPX formation (starting material), it is obvious that residual water impacts the $NaBH_4$ (reduction agent) mediated reduction step and thus the particle size distribution. Therefore, a detailed description is provided in the experimental section on how to successfully carry out a quasi water-free phase transfer. After the reaction ended the suspension was purified through DCM/hexane redissolution/precipitation for multiple times and extracted with methanol (abbreviated as DCM/Hex/MeOH), resulting in an optical spectrum of the "purified" $Au_{25}$ NC with sharp, molecule-like optical transitions at 415 and 670 nm, shown in black in Fig. 1A.

The present method is designed to be lossless and convenient in two ways: First, purification typically employed before thiol-etching is instead performed afterwards, ensuring a larger fraction of $Au_n$-TPP NPs to be loss-free and available for the conversion to $Au_{25}$ NCs. Second, an adjustment of the Au:thiol molar ratio through thermo gravimetric analysis (including drying and combustion of 5-10 mg NPs) can be bypassed, and the suggested thiol volume can be considered as constant. Combining an optimized reaction time derived from the optical spectrum, proper handling of water residues, and finally a loss-free transfer of polydisperse Au NPs by omitting purification in between of synthesis steps, the synthetic yield of "purified" $Au_{25}$ NCs has been raised to 35 mg NCs per 100 mg chloroauric acid (40% in respect to elemental Au).

Various analytic and imaging techniques were applied to further investigate the properties and quality of the $Au_{25}$ NC suspension. Transmission electron microscopy (TEM) and high resolution TEM (shown in Fig. 1B and inset) reveal well separated nm sized crystals, with no aggregates or larger particles observed. By means of high-resolution electrospray ionization (ESI) MS, shown in Fig. 1C, three modifications of the $[Au_{25}(TPP)_{10}(PET)_5X_2]^{2+}$ NC are identified, giving rise to signals at m/z 4151.60, 4174.08 and 4196.05, corresponding to $X_2=Cl_2$, BrCl or $Br_2$ respectively. The solubility of $Au_{25}$ NCs in solvents of varying polarities, from methanol to toluene, makes their separation from residual organic synthesis byproducts challenging. This topic is not covered sufficiently in previous reports, which lack $^{31}P$ or $^1H$ NMR spectra. [13 or 14, 23–25] Through these methods, we identified a large amount of TPP and the ionic and organic soluble phase transfer agent tetraoctylammonium ($TOA^+$) as major organic residues. The DCM/Hex/MeOH purification sequence thoroughly removed loose TPP, resulting in exclusively surface-bound TPP giving rise to a signal at 53 ppm in the $^{31}$P NMR spectrum (Fig. 1D). In addition, a remarkably low concentration of residual TOA$^+$, identified by the small integral of the signal at 3.15 ppm in the $^1$H NMR spectrum (Fig. S1) is achieved.

Since organic byproducts were thoroughly removed as evidenced by the NMR analysis, a deeper insight into colloidal impurities of inorganic nature was necessary. Their presence can be clearly recognized from the UV-Vis optical spectra. Comparing the UV-Vis optical spectra of Au$_{25}$ NC suspensions prepared by different groups a remarkable difference is evident (Fig. 2). The extinction offset compared to the black curves is clearly visible below 460 nm. An offset in this spectral region could originate from NCs of a similar or smaller core size than Au$_{25}$. However, distinct spectral signatures of such NCs were not apparent in the corresponding optical or mass spectra. Consequently, plasmonic absorption or scattering by large Au NPs cause the increased extinction and broaden the 415 nm optical feature of the Au$_{25}$ NC. Removal of these by the established DCM/Hex/MeOH purification sequence and especially proper reaction control, including an optimized reaction time, were identified in the present work to be crucial for obtaining a Au$_{25}$ NC suspension free of large NPs and an extinction spectrum as shown in black in Fig. 2.

As apparent from the blue curve in Fig. 2, an extinction offset to the black curve can also rise only below 380 nm. This indicates that instead of Au NPs, unreacted starting material AuTPPX[13] or, despite their absence in the reported MS spectrum, small few gold-atoms comprising complexes are present. To this day, mass spectra of products from thiol etching synthesis as applied for Au$_{25}$ and for many other NCs are usually not shown below m/z 1200, although the formation of byproducts including few gold-atoms containing complexes is known giving rise to signals in that mass range. A 2011 report from the Jin group shed light on the dark: The Au$_2$ complex with the formula [Au$_2$TPP$_2$PET]$^+$ was crystallized from a Au$_{25}$ NC suspension.[15] Although crystals of this major byproduct could be mechanically separated, it did not become clear whether a significant level of purity of the Au$_{25}$ NC suspension could be achieved since a mass spectrum covering the whole mass range was not reported. Therefore, the present study focused on a comprehensive MS analyses including the mass range below m/z 1200 and removal of all inorganic byproducts, including the few gold-atoms comprising complexes.

High performance liquid chromatography (HPLC) proved to be a useful tool for a preparative separation of Au NCs covered with different ligands and of varying size.[26,27] Combined with high resolution electrospray ionization (ESI)-MS, a chromatographic separation of the $Au_2$ complex from $Au_{25}$ NCs was targeted. With the established HPLC method, three distinct fractions could be observed, as indicated by the base peak chromatogram (BPC) in Fig. 3A. The first fraction solely includes the above mentioned $[Au_2TPP_2PET]^+$ complex (m/z 1055) followed surprisingly by a third single positive charged compound with a mass of m/z 1409, whose simulated spectrum is shown in the inset. After about 25 min, all three modifications (with $X_2$= $Cl_2$, ClBr or $Br_2$) of the $Au_{25}$ NC are eluted, along with a compound already present in the blank measurement. The corresponding BPC of the blank measurement as well as extracted ion chromatograms (EICs) are shown in Fig. S2.

Since a gold-complex with a mass of 1409 has not yet been reported, matrix assisted laser desorption ionization (MALDI)-MS was applied to analyze the peak at m/z 1409 in more detail. A MALDI-MS spectrum in the range of m/z 500-5040 of the $Au_{25}$ NC sample is shown in Fig. 3B. In accordance to the HPLC-ESI-MS results, the compounds with masses of 1409 and 1055 give rise to a signal and a fragment at m/z 721. All three compounds were further fragmented and analyzed by means of MALDI-MS/MS (Fig. 3C). Indicating a similar molecular structure, the compounds at m/z 1409 and 1055 fragment in a similar way to fragments at m/z 721 and 458, which are suggested to be $[AuTPP_2]$ and $[AuTPP]$, respectively. In addition, the MS/MS spectrum of the peak at m/z 1409.2 compound shows a fragment at m/z 1147, which can be assigned to $[Au_3TPP_2O_2]$. Based on this fragment and the well matching simulated spectrum (inset in Fig. 3A) the newly found compound at m/z 1409 is proposed to be a $[Au_3TPP_3O_2]$ complex, which has not been reported to this day.

Having identified the residual inorganic byproducts including the $Au_2$ and $Au_3$ complexes, we proceeded to remove them. In addition to the chromatographic separation, two practical methods are presented to obtain Au$_{25}$ NCs satisfying different application purposes. The first method is based on a known procedure and involves the crystallization of the Au$_{25}$ NCs whose microscope image is shown in Fig. 4B. However, in accordance with a previous report[15], a mass spectrum indicating the exclusive presence of Au$_{25}$ NCs could not be obtained from the redissolved crystals. Conclusively, the mechanical separation of Au$_2$/Au$_3$ complexes from the Au$_{25}$ appeared to be useful only when few single crystals are required (e.g. single crystal analysis) but not as a purification method for the suspension.

The second method is based on a classical selective precipitation method. Starting with the "purified" Au$_{25}$ NC suspension with an UV-Vis optical spectrum shown in Fig. 4A in black, the dry material is mixed with a methanol/water (2/8 vol.%) mixture resulting a colorless liquid phase and a residual black solid. The optical spectrum of the liquid phase is shown in red in Fig. 4A. Two of four sharp optical features, located at 266 and 273 nm are directly assigned to the [Au$_2$TPP$_2$PET]$^+$ complex[15] whereas two residual features at 254 and 260 nm are believed to stem from absorption of the [Au$_3$TPP$_3$O$_2$] complex. The residual black solid is redissolved in methanol resulting in an optical spectrum, shown in blue in Fig. 4A, indicative of Au$_{25}$ NCs. The spectrum of the obtained Au$_{25}$ NCs gives rise to two new sharp transitions and rises moderately below 300 nm, while the starting materials and Au$_2$/Au$_3$ spectra rise much steeper. As the black spectrum appears to be a superposition of the red Au$_2$/Au$_3$ spectrum and the blue Au$_{25}$ spectrum (indicated by the inset), the UV-Vis spectral data suggests that Au$_{25}$ NCs are successfully separated by selective precipitation from its major byproducts. The results from the UV-Vis spectroscopy are supported by ESI-MS recorded before and after selective precipitation in the range of m/z 800 to 5000, shown in Fig. 4C. It becomes clear that "purified" Au$_{25}$ NCs are present in vanishingly small quantities compared to the Au$_2$ complex. However, after selective precipitation, all aforementioned byproducts are removed and solely Au$_{25}$ NCs remain in the suspension.

## Conclusion

The present work demonstrates the necessity of monitoring the Au nanoparticle etching process by UV-VIS spectroscopy, and the importance of proper purification and selective precipitation to obtain ultra-pure $Au_{25}$ NCs. The above spectra show that despite the classically applied and herein advanced DCM/Hex/MeOH purification techniques, $Au_{25}$ NCs prepared by both, the present and previous methods mainly contain the $Au_2$ complex. The key component was a thorough MS analysis, including a discussion of the mass region below m/z 1200, which was not covered in previous reports. An HPLC method was established, which serves as a starting point for a preparative chromatographic separation leading to ultra-pure $Au_{25}$ NC fractions. The preparative HPLC and selective precipitation methods established here, are believed to be directly applicable to the purification of other Au NCs. Similar byproducts might occur in the synthesis of highly fluorescent $Ag_xAu_{25-x}$, single atom-doped $Au_{24}Pd$, $Au_{25}$ NCs' big brother $Au_{37}$, and many other mix-ligand stabilized charged Au NCs. Due to the size focusing mechanism, the formation of these small gold-containing byproducts is unavoidable.[28] Their removal, along with that of organic residues, is key to obtaining high purity required for using Au NCs in applications.

## Acknowledgments

The authors thank the German Research Foundation DFG financial support in the frame of the Cluster of Excellence "Center of ultrafast imaging CUI" and for granting the project KL 1453/9-2. The European Research Council is acknowledged for funding an ERC Starting Grant (Project: 2D-SYNETRA (304980), Seventh Framework Program FP7). Special thanks goes to Andreas Kornowski and Stefan Werner for transmission electron microscopy.

**Figures**

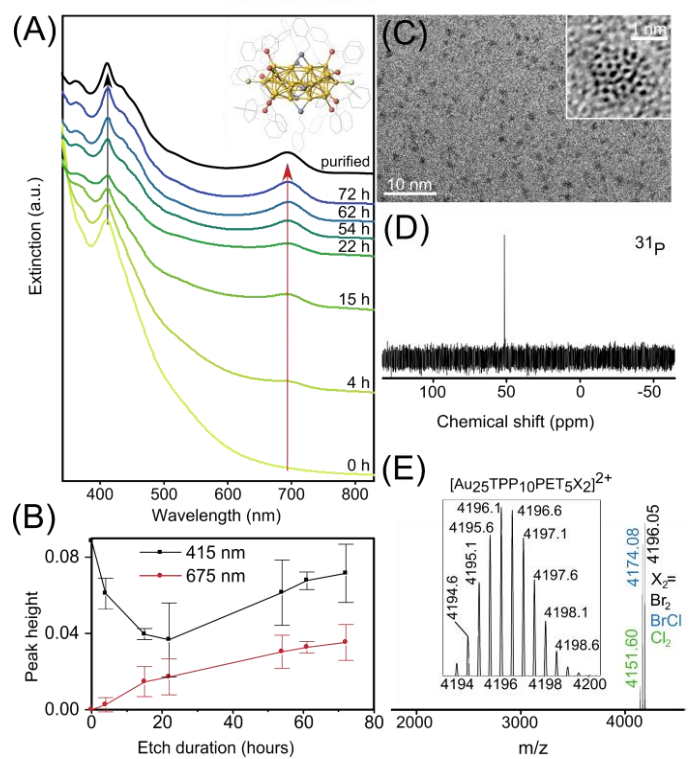

Fig. 1: A) Conversion of polydisperse Au NPs into $[Au_{25}(TPP)_{10}(PET)_5X_2]^{2+}$ NCs followed by UV-Vis spectroscopy. The crystal structure is shown in the inset. Inset color labels: Au, yellow; S, blue; P, red; X: Cl and/or Br, green. B) Evolution of the peak height at 415 and 675 nm during the etching process. Data and error bars were obtained by averaging data sets from the syntheses and plotting the standard deviation. C) TEM and HRTEM (inset) of drop casted $Au_{25}$ NCs. D) $^{31}P$ NMR spectrum of $Au_{25}$ NCs in $CD_2CL_2$. E) High resolution ESI-MS of $Au_{25}$ NCs and zoom into the region of m/z 4196 (inset).

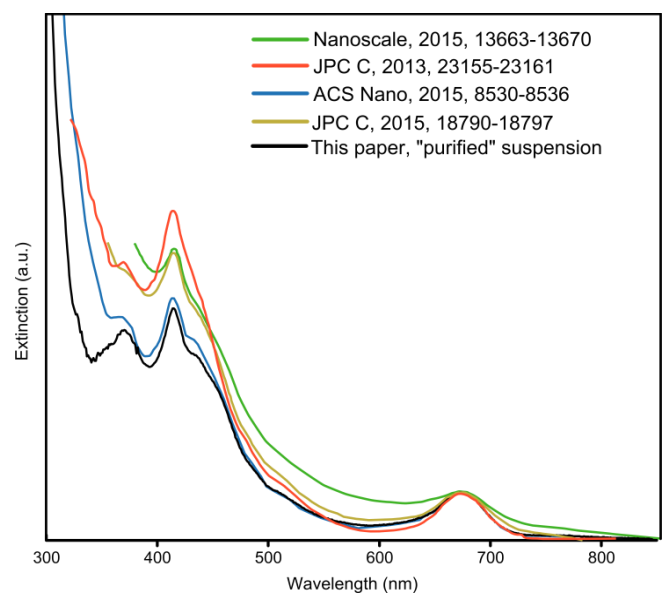

Fig. 2: Optical spectra reported for $[Au_{25}(TPP)_{10}(PET)_5X_2]^{2+}$ NCs illustrating clear differences in the relative intensities of the peaks and the expression of the fine structure. All data were digitized from the cited source and normalized to 675 nm (HOMO to LUMO transition).

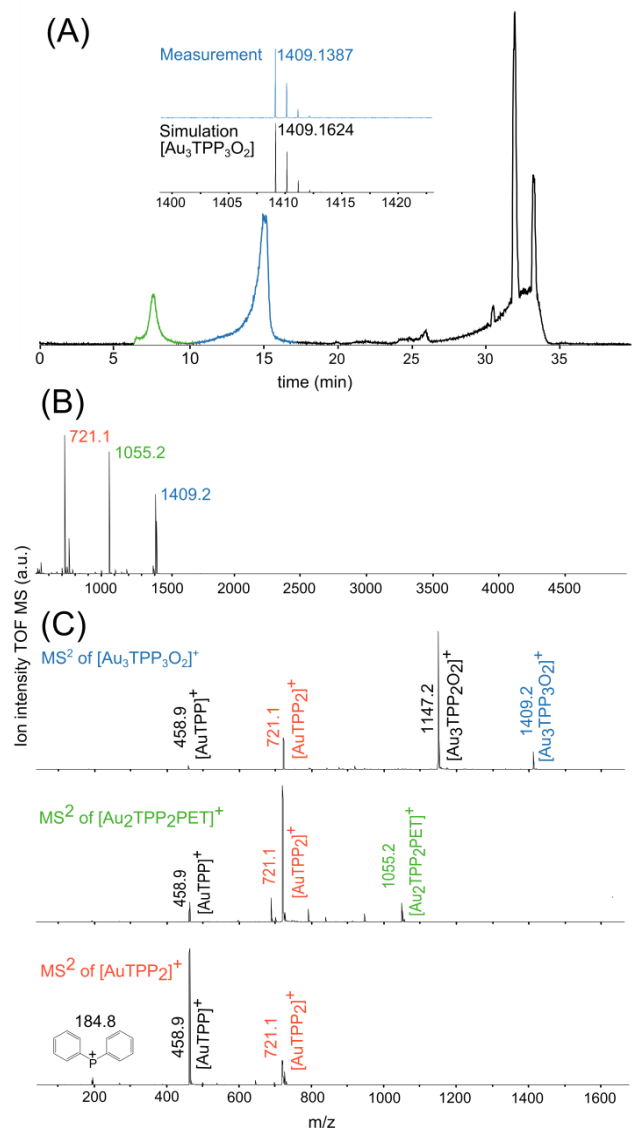

Fig. 3: MS analysis of $[Au_{25}(TPP)_{10}(PET)_5X_2]^{2+}$ and contained byproducts. A) Base peak chromatogram of $Au_{25}$ NCs measured with ESI-MS. B) MALDI-MS of $Au_{25}$ NCs. C) MALDI-MS/MS analysis of components present in B). Same color in different figures indicates the same compound or the same fragment.

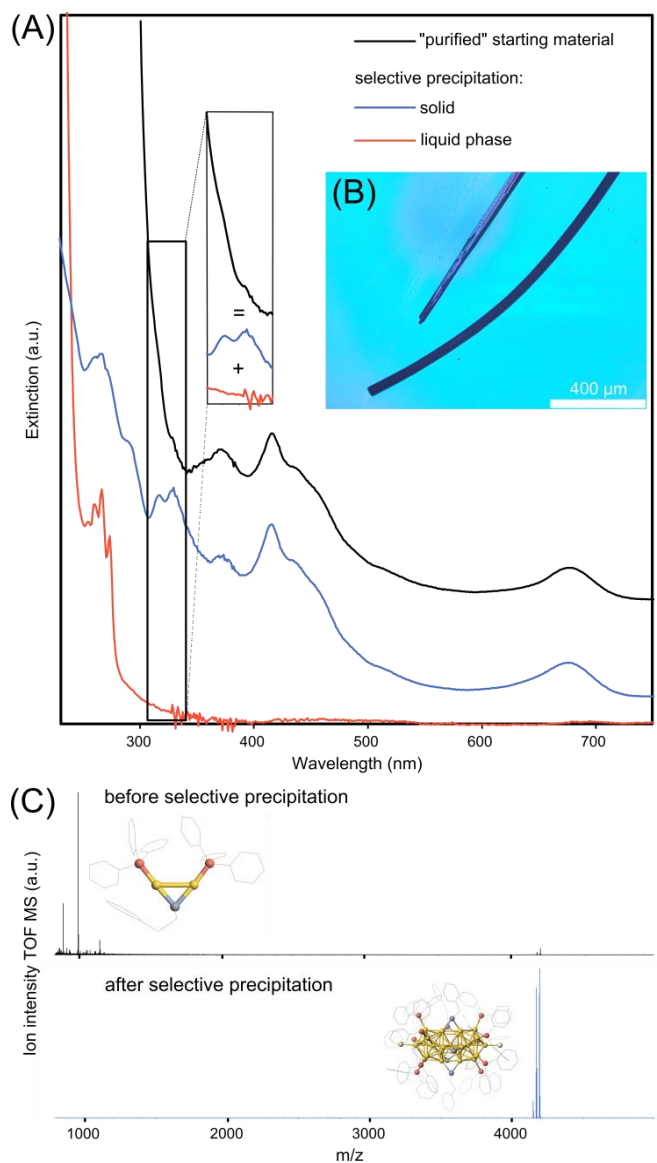

Fig. 4: A) UV-Vis spectra of the "purified" $[Au_{25}(TPP)_{10}(PET)_5X_2]^{2+}$ NC suspension and from it selectively separated compounds, y-offset for a better overview. The inset illustrates the broadening of $Au_{25}$ NCs optical features by the presence of $Au_2/Au_3$ complexes. B) Optical microscope image of $Au_{25}$ NC crystals. C) ESI-MS analysis of "purified" $Au_{25}$ NC suspension and after selective purification. The inset shows the crystal structure of the most intensive component, $Au_2$ complex and $Au_{25}$ NC, respectively.